# Carbon Nanotube Terahertz Polarizer

Lei Ren,[1,2] Cary L. Pint,[2,3,4] Layla G. Booshehri,[1,2] William D. Rice,[1,2,3] Xiangfeng Wang,[1,2,3,$]

David J. Hilton,[1,2,†] Kei Takeya,[5] Iwao Kawayama,[5] Masayoshi Tonouchi,[5] Robert H. Hauge,[2,4] and

Junichiro Kono[1,2,3,*]

[1]Department of Electrical and Computer Engineering, [2]The Richard E. Smalley Institute for Nanoscale Science and Technology, [3]Department of Physics and Astronomy, [4]Department of Chemistry, Rice University, Houston, Texas 77005, USA.  [5]Institute of Laser Engineering, Osaka University, Yamadaoka 2-6, Suita, Osaka 565-0871, Japan

*Correspondence and requests for materials should be addressed to Junichiro Kono (kono@rice.edu).

[$]Present address: Intelligent Automation, Inc., Rockville, Maryland 20855.

[†]Present address: Department of Physics, University of Alabama, Birmingham, Alabama 35294.

**We describe a film of highly-aligned single-walled carbon nanotubes that acts as an excellent terahertz linear polarizer. There is virtually no attenuation (strong absorption) when the terahertz polarization is perpendicular (parallel) to the nanotube axis. From the data we calculated the reduced linear dichrosim to be 3, corresponding to a nematic order parameter of 1, which demonstrates nearly perfect alignment as well as intrinsically anisotropic terahertz response of single-walled carbon nanotubes in the film.**

The one-dimensional character of confined carriers and phonons in single-walled carbon nanotubes (SWNTs) leads to extremely anisotropic electric, magnetic, and optical properties.[1] Individual metallic SWNTs have been shown to be excellent quantum wires with very long coherence lengths,[2] while individual semiconducting SWNTs have been shown to strongly absorb and emit light only when the light polarization is parallel to the tube axis.[3] Moreover, individualized SWNTs, both metallic and semiconducting, have been shown to align well in an external magnetic field,[4-9] due to their anisotropic magnetic susceptibilities,[10-12] and aligned SWNTs exhibit strong linear dichroism due to their anisotropic optical properties. Even in bundled samples, where SWNTs form aggregates through van der Waals attraction, their anisotropic properties are expected to be preserved to a large degree, even though systematic optical spectroscopy studies have been limited[13-15] due to the rarity of ensemble samples in which the SWNTs are highly aligned.

Here we report results of terahertz (THz) transmission measurements on a strongly absorbing film of highly aligned SWNTs, which demonstrate extremely high anisotropy. Strikingly, when the THz polarization was perpendicular to the alignment axis, no absorption was observed within our experimental sensitivity despite the macroscopic thickness (~2 μm) of the film; when the polarization was parallel to the alignment direction, there was strong absorption. The degree of polarization in terms of absorbance was 1 and the reduced linear dichroism was 3, throughout the entire frequency range of our experiment (0.1-1.8 THz). Using the theory of linear dichroism for an ensemble of anisotropic molecules,[16] we show that this value of reduced dichroism (i.e., 3) is possible only when the nematic order parameter ($S$) is 1. These observations are a direct result of the one-dimensional nature of conduction electrons in the nanotubes and, at the same time, demonstrate that any misalignment of nanotubes in the film must have characteristic length scales much smaller than the wavelengths used in these experiments (1.5 mm – 150 μm). All these findings suggest that this type of aligned SWNT film performs as an ideal linear polarizer in the THz frequency range.

In order to produce the aligned SWNTs for this study, 2 μm wide pads of catalyst containing 0.5 nm Fe and 10 nm $Al_2O_3$, having spacing of 50 μm between lines, were formed on a Si wafer by using optical lithography and electron-beam deposition, as recently described elsewhere.[17] The catalyst pads were then grown in a hot-filament chemical vapor deposition apparatus maintained at 750°C and 1.4 Torr utilizing a water-assisted growth process with $C_2H_2$ decomposition.[18-20] The as-grown lines of aligned SWNTs initially adopt a vertical orientation with respect to the growth substrate, with a length that can be determined based on the duration of catalyst exposure to the growth conditions. Following the growth process, a high temperature (750°C) $H_2O$ vapor etch is employed to free the catalyst-SWNT interface,[17] allowing efficient transfer of an aligned film to a host substrate of choice; sapphire in this case. The transfer method utilized in this study relies on the simple concept of strong side-wall van der Waals adhesion between SWNTs and the transfer substrate, and weaker SWNT-end interaction with the growth substrate following SWNT-catalyst interface etching. This is analogous to the mechanism behind the "Gecko" effect.[21] The result of the transfer process is a homogenous film (initially ~2 μm thick) that remains as-grown, highly aligned, and free of exposure to any sort of solvent or liquid. Figure 1a shows a top-down scanning electron microscope image of the SWNT alignment present in such a transferred film, emphasizing the high degree of alignment, which makes this film well-suited for the study presented in this work.

The THz setup we utilized to study this SWNT sample was a typical time-domain THz spectroscopy system based on photoconductive antennas, where both the THz emitter and detector were made of low-temperature grown GaAs.[22] The THz beam obtained from such an emitter was already highly linearly polarized, but a free standing wire-grid polarizer with a degree of polarization of more than 99.5% in this range was placed in the path of the incident THz beam (8 inches both from the emitter and sample) to ensure the high degree of polarization of the THz beam incident on the sample. As schematically shown in Fig. 1b, the SWNT sample was rotated about the propagation direction of the THz wave, which changed the angle, $\theta$, between the nanotube axis and the THz electric field polarization direction from 0° to 90°. Thus, polarization-dependent THz transmission measurements were performed on both the SWNT film sample on a sapphire substrate and a reference sapphire sample with the same thickness as the sample substrate.

Figure 2a shows time domain waveforms for THz waves transmitted through the reference sapphire (black dashed curve) and the SWNT film (solid colored curves). The different colors represent different angles $\theta$ between the THz polarization direction and the nanotube alignment direction. For the $\theta = 90º$ case, the signal was essentially the same as that of the reference, which means that *there was virtually no interaction when the nanotube axis was perpendicular to the THz electric field*. For the $\theta = 0º$ case, the signal was the smallest, which means that, when the nanotube axis was parallel with the THz electric field, the interaction was the strongest. From the $\theta = 45º$ and 30º cases, we see that the transmitted THz signal decreases monotonically as $\theta$ decreases.

Figure 2b shows the amplitude spectra after Fourier transformation of the time domain data in Fig. 2a in a range of 0.2-1.8 THz, where the anisotropy is more obvious. Figure 2c shows the absorbance, $A = -\text{Log}_{10}(T)$, where $T$ is the transmittance defined as $T = |E_s/E_r|^2$ and $E_s$ and $E_r$ are the complex THz signals in the frequency domain obtained through Fourier transform of the time domain data for the sample and reference, respectively. From this figure, we can clearly see that as the angle $\theta$ increases from 0º to 90º, the absorbance of the SWNTs decreases monotonically. When $\theta = 90º$, the absorbance is zero throughout this frequency range. On the other hand, when $\theta = 0º$, the absorbance is finite and high; it increases with increasing frequency, reaching a value over 1.0 at 1.8 THz. The 30º and 45º absorbance lines show the same trend as the 0º curve but with smaller amplitudes. The increasing absorbance with frequency in this spectral range is consistent with previous far-infrared spectroscopy results on various types of SWNT samples showing a robust absorption peak around 4 THz, whose origin is not understood.[23-26]

To quantify the degree of alignment of SWNTs in this film from these THz transmission data, we employ a data analysis procedure developed for studying anisotropic optical properties of SWNTs in the optical range.[4-9] Figure 3a plots the parallel ($A_{//}$) and perpendicular ($A_\perp$) absorbance spectra, corresponding to $\theta = 0º$ and 90º, respectively, together with $A_0 = (A_{//} + 2A_\perp)/3$, the isotropic absorbance.[16] This physical quantity represents the absorbance expected if the nanotubes were randomly oriented. Finite alignment moves up (down) $A_{//}$ ($A_\perp$) with respect to $A_0$ and induces a finite linear dichroism, $LD = A_{//} - A_{\perp}$, shown in Fig. 3b. However, it is the *reduced* linear dichroism, $LD^r \equiv LD/A_0$, that provides a normalized measure of alignment. For example, $LD$ increases with the film thickness, while $LD^r$ remains the same, as we have shown in different samples (data not shown). From a microscopic viewpoint, the $LD^r$ can be expressed as

$$LD^r = 3[(3\cos^2\alpha - 1)/2] \cdot S, \qquad (1)$$

where $\alpha$ is the angle between the long axis and the direction of the dipole moment and $S$ is the nematic order parameter (= 0 when the nanotubes are randomly oriented and = 1 when the nanotubes are perfectly aligned).[16] Figure 3b indicates that within our experimental range $LD^r$ is nearly constant at 3, which, combined with Eq. (1), indicates that $S \sim 1$, assuming that $\alpha = 0$. Note that any finite value of $\alpha$ would result in a value of $S$ that is larger than 1, which is impossible by definition. Therefore, *the fact that $LD^r \sim 3$ proves not only that the nanotubes are well aligned ($S \sim 1$) but also that the THz response of SWNTs is intrinsically anisotropic ($\alpha \sim 0$)*. Finally, as an important parameter for a polarizer,[27] we calculated the degree of polarization, i.e., $P = (A_{//} - A_{\perp})/(A_{//} + A_{\perp})$, which is plotted in Fig. 3c. As shown, it is very close to 1 throughout the entire region. All of this data suggests that this THz polarizer is of excellent quality.

In summary, we have synthesized an optical polarizer made of highly-aligned single-walled carbon nanotubes on a sapphire substrate, which works very well in the THz frequency range. Very anisotropic THz absorbance data demonstrates the extremely high degree of alignment of the SWNTs in the film. The measured degree of absorbance polarization is 1, while the measured reduced linear dichroism is 3, which corresponds to an order parameter of 1, demonstrating that any nanotube misalignment must have length scales much smaller than the wavelengths used (1.5 mm – 150 μm). Additionally, our data clearly show that THz response is virtually non-existent when the THz field is perpendicular to the nanotube axis, and this intrinsic property is present even in a bulk sample in which the nanotubes are bundled, as long as they are well aligned.

This work was supported by the Department of Energy (through Grant No. DE-FG02-06ER46308), the National Science Foundation (through Grant No. OISE-0530220), and the Robert A. Welch Foundation (through Grant No. C-1509). The authors thank Dr. Daniel Mittleman for helpful discussions.

**Supporting Information Available.** Detailed experimental methods. This material is available free of charge via Internet at http//pubs.acs.org.

**Figure 1** **(a)** Scanning electron microscope image of the highly horizontally aligned SWNT film used in this experiment. **(b)** Sketch of the experimental configuration used, showing the interaction between the linearly polarized THz electric field and highly aligned SWNT film. The angle between the THz polarization direction and the nanotube alignment direction, $\theta$, was varied between 0º and 90º.

**Figure 2** **(a)** THz electric field signals in the time domain, **(b)** transmitted THz amplitude spectra obtained through Fourier transform of the time domain signals in (a), and **(c)** THz absorbance spectra, for the reference sapphire substrate (black dashed curves) and for the SWNT film for different polarization angles (colored solid curves).

**Figure 3** **(a)** Parallel ($A_{//}$), perpendicular ($A_\perp$), and isotropic ($A_0$) absorbance, **(b)** linear dichroism (*LD*) and reduced linear dichroism (*LD$^r$*), and **(c)** degree of polarization (DOP) as a function of frequency, measured for the SWNT film.

(a)

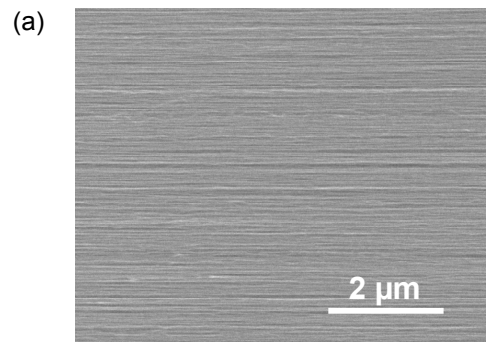

(b)

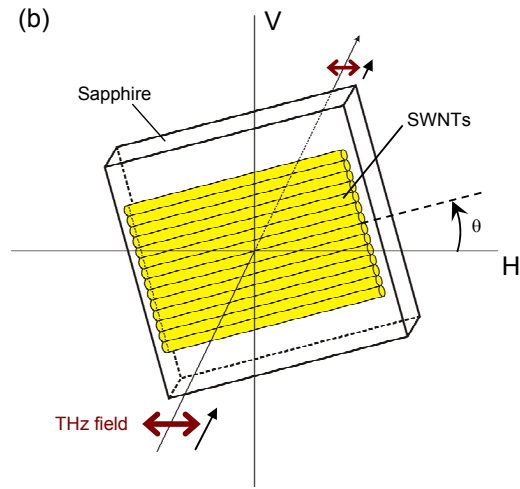

Figure 1

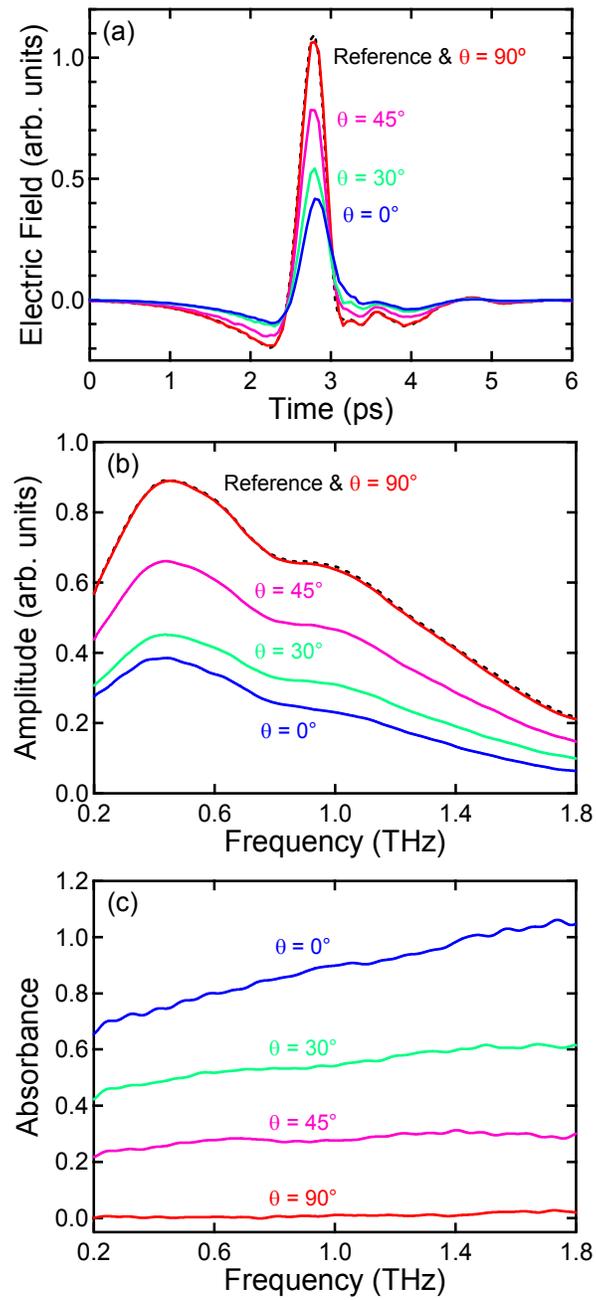

Figure 2

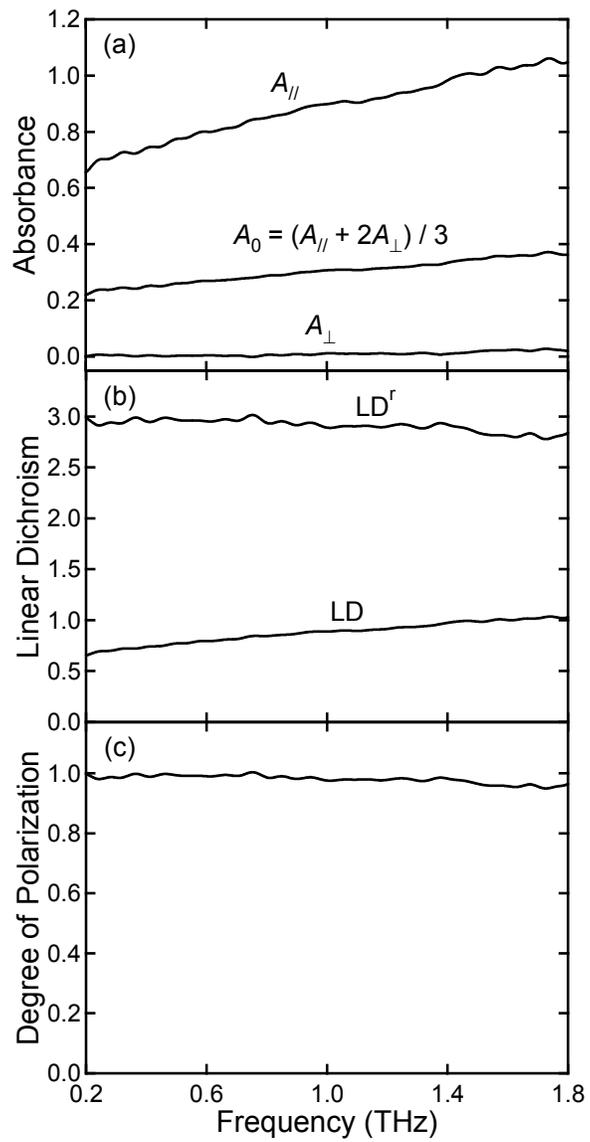

Figure 3